\documentclass[11pt,a4paper]{article}
\usepackage[utf8]{inputenc}
\usepackage{amsmath,amssymb}
\usepackage[english]{babel}
\usepackage{graphicx}
\usepackage{wrapfig}
\usepackage{url}
\usepackage{graphicx}
\usepackage{tabularx}
\usepackage[font={small,it}]{caption}
\usepackage{listings}
\usepackage{scalerel}
\usepackage{calligra}
\DeclareMathAlphabet{\mathcalligra}{T1}{calligra}{m}{n}
\usepackage{float}
\usepackage{amsfonts}
\usepackage{xcolor}
\usepackage{physics}
\usepackage{ulem}
\usepackage{hyperref}
\usepackage{mathtools}
\usepackage{tcolorbox}
\usepackage{soul,xcolor}
\usepackage[toc,page]{appendix}
\usepackage{a4wide}
\usepackage{csquotes}

\usepackage{pdflscape}
\usepackage{parskip}
\usepackage{geometry}
\newenvironment{aleq}
    {\begin{equation}\begin{aligned}}
    {\end{aligned}\end{equation}\ignorespacesafterend}
\usepackage{subfiles}
\graphicspath{{figures/}}
\usepackage{subcaption}
  \usepackage{a4wide}
  \usepackage{latexsym}
  \usepackage{epsf}
  \usepackage{amssymb}
  \usepackage{graphicx}
  \usepackage{amsmath, cite}
  \usepackage{amsmath,amssymb,amsthm}
  \usepackage{verbatim}
  \usepackage{hyperref}
  \usepackage{color}
  \usepackage{mathtools}
  \usepackage{soul,xcolor}

\usepackage{makecell}

\newcommand{\be}{\begin{equation}}
\newcommand{\ee}{\end{equation}}

\begin{document}
\numberwithin{equation}{section}

\vspace{2.718cm}
\begin{center}

{\LARGE \bf{
On the DGKT brane dual and its decoupling}}\\

\vspace{1cm}

 {\large Fien Apers}\\
 \vspace{0.5 cm}
 {\small Instituto de F\'{i}sica Te\'{o}rica IFT-UAM/CSIC,
C/ Nicol\'{a}s Cabrera 13-15, Campus de Cantoblanco, 28049 Madrid, Spain}\\
\vspace{0.5 cm} {\small\slshape fien.apers@uam.es}\\

\vspace{1cm}

{\bf Abstract} \end{center} {
It is not understood whether scale-separated AdS vacua in string theory admit a holographic dual. A well-known class of such vacua is provided by the DGKT solutions of massive type IIA string theory, where scale separation arises from large fluxes. In this work, we construct a ten-dimensional brane geometry whose near-horizon limit reproduces the DGKT vacua, using a flux-backtracking approach combined with intersecting D4-brane stacks dual to the unbounded flux sector.

We then use this setup to test whether the brane worldvolume theory decouples from the bulk. Modes localised near the branes, deep in the AdS throat, are found to be infinitely redshifted with respect to asymptotic observers. Moreover, an analysis of graviton fluctuations shows the presence of an infinite potential barrier near the branes, providing a direct indication of decoupling. We conclude by comparing these results with recent arguments against decoupling in scale-separated AdS vacua, which focus on the asymptotic region where modes are blueshifted.}

\newpage

\tableofcontents
\section{Introduction}
The existence of holographic duals of scale-separated $\mathrm{AdS}$ vacua (see \cite{Coudarchet:2023mfs} for a review) is an important open question in string theory. Scale-separated $\mathrm{AdS}$ vacua exhibit a hierarchy between the Hubble scale and the Kaluza--Klein scale of the extra dimensions. This hierarchy is required for a reliable lower-dimensional effective description, and such vacua are often a first step toward constructing de Sitter solutions. Known $\mathrm{AdS/CFT}$ examples \cite{Maldacena:1997re, Aharony:1999ti} do not display this hierarchy, raising the question of whether holographic duals in string theory necessarily involve large extra dimensions. On the conformal field theory side, scale separation would correspond to a large gap in the spectrum of single-trace primary operators, and such CFTs are currently unknown \cite{Collins:2022nux}. This leaves two intriguing possibilities: either there is a general holographic obstruction to achieving scale separation (for work in this direction, see \cite{Bobev:2023dwx, Perlmutter:2024noo, Alday:2019qrf}), or there exists an entirely novel class of holographic CFTs.

A well-known class of scale-separated $\mathrm{AdS}_4$ vacua are the DGKT--CFI \cite{DeWolfe:2005uu, Camara:2005dc} vacua in massive type IIA string theory. Scale separation is obtained by increasing an $F_4$ flux that is not bounded by tadpole constraints. These vacua are appealing from a holographic point of view because the large flux hints at a straightforward large-$N$ structure, similar in spirit to familiar Freund--Rubin setups. Other related $\mathrm{AdS}_4$ and $\mathrm{AdS}_3$  vacua with such a large-$N$ structure include \cite{Cribiori:2021djm, VanHemelryck:2022ynr, VanHemelryck:2025qok, Farakos:2020phe,Farakos:2025bwf, Arboleya:2024vnp, Arboleya:2025lwu, Arboleya:2025jko, Carrasco:2023hta}.

Earlier holographic studies \cite{Aharony:2008wz, Conlon:2021cjk, Apers:2022tfm, Apers:2022vfp, Apers:2022zjx, Bobev:2025yxp} have already uncovered several interesting features for these vacua. The conformal dimensions of operators dual to stabilised moduli approach integer values in the large-$N$ limit, accompanied by spacetime-dependent shift symmetries on the AdS side that are not yet fully understood \cite{Apers:2022vfp, Bonifacio:2018zex, Blauvelt:2022wwa}. The holographic central charge scales as $c \sim N^{9/2}$, which is unusually large and suggests that a rather special brane construction may be required to account for it.

In Section~2, we construct a metric that reproduces the DGKT vacua in its near-horizon limit. Our approach consists of two steps. First, we apply the flux backtracking method introduced in~\cite{Apers:2025pon} to obtain a background to be probed by the branes carrying the DGKT dual theory. This amounts to solving BPS flow equations driven by the fluxes that do not scale with $N$: the Romans mass and the $H_3$ fluxes. Second, we dualise the unbounded $F_4$ fluxes to three stacks of intersecting D4-branes and place these stacks on the backtracked background. We verify that the resulting ten-dimensional metric solves the Einstein and dilaton equations and contains $\mathrm{AdS}$ vacua in its near-horizon with the same properties as the original DGKT solutions. In particular, the three intersecting stacks on this strongly coupled singularity appear to account for the unusually large scaling of the holographic central charge.

As an application, we use this geometry in Section~3 to test whether the brane worldvolume theory decouples from the bulk. We show that modes localised near the branes deep in the AdS throat are infinitely redshifted relative to asymptotic observers. As a direct probe of decoupling, we compute the effective Schrödinger potential for gravitational waves (and scalar modes) and demonstrate the emergence of an infinite potential barrier near the location of the branes: a signature of decoupling between brane and bulk degrees of freedom. We conclude with a discussion comparing these results with recent claims in the literature \cite{Bedroya:2025ltj}, where the absence of decoupling in scale-separated AdS vacua was inferred from the behaviour of the asymptotic region.

\section{Construction of the DGKT brane theory}

In this section, we construct a ten-dimensional metric for a brane geometry that approaches the DGKT vacua in the near-horizon limit.  

The DGKT vacua arise from compactifying massive type IIA string theory on a Calabi--Yau manifold in the presence of $F_4$, $H_3$, and $F_0$ fluxes together with orientifold planes. The resulting solutions are four-dimensional $\mathrm{AdS}$ vacua with minimal supersymmetry and all moduli stabilised. The $F_4$ flux is unbounded by the orientifold tadpole and generates both parametric control and scale separation.

In terms of the universal moduli, namely the volume modulus and the four-dimensional dilaton,
\begin{align}
    u = \mathrm{vol}_S^{1/3}, \qquad s = e^{-\phi}\sqrt{\mathrm{vol}_S},
\end{align}
where $\mathrm{vol}_S$ denotes the internal volume in string frame, the effective potential takes the form
\begin{equation}\label{DGKT}
    V = \frac{1}{s^3}\left[
    \frac{A_{F_4}}{u s}
    + \frac{A_{F_0} u^3}{s}
    + \frac{A_{H_3} s}{u^3}
    - A_{O6}
    \right],
\end{equation}
with coefficients proportional to quantised flux squares,
\begin{align}
    A_{F_4} \sim N^2, \qquad
    A_{H_3} \sim h^2, \qquad
    A_{F_0} \sim m^2,
\end{align}
and $A_{O6} = 4\sqrt{A_{F_0} A_{H_3}}$. At the $\mathrm{AdS}$ minimum, the moduli take the values
\begin{align}\label{DGKTprop1}
    \mathrm{vol}_S \sim N^{3/2} m^{-3/2}, 
    \qquad
    e^\phi \sim N^{-3/4} m^{-1/4} h,
\end{align}
and the potential energy is
\begin{align}\label{DGKTprop2}
    V_{\mathrm{min}} \sim - \frac{h^4 m^{5/2}}{N^{9/2}}.
\end{align}
The masses of the stabilised moduli satisfy
\begin{align}
    m^2 R_{\mathrm{AdS}}^2 = 70, \quad 18,
\end{align}
which correspond to the surprisingly simple integer conformal dimensions $10$ and $6$ of the dual scalar operators; see~\cite{Apers:2022tfm, Apers:2022vfp, Plauschinn:2022ztd, Conlon:2021cjk} for a discussion.

\subsection{Flux backtracking to find the singularity}

To find the background geometry, we first perform flux backtracking with non-zero Romans mass $m$ and $h$ units of $H_3$ flux, as in \cite{Apers:2025pon}. For simplicity, we take the internal manifold to be toroidal.

A solution to the BPS flow equations is given by
\begin{equation}
u(r) \sim r^{2/9}, 
\qquad 
s(r) \sim r^{2/3}, 
\qquad 
A(r) = \frac{13}{27}\ln r,
\end{equation}
where $u=\mathrm{vol}_S^{1/3}$ is the volume modulus, $e^\phi$ is the four-dimensional dilaton, and the four-dimensional metric has the form $ds_4^2 = dr^2 + e^{2A(r)} ds_{3,M}^2$.

This leads to the following ten-dimensional geometry, 
\begin{align}\label{BG}
ds_{10}^2 &= dr^2
+ r^{-10/9} ds^2_{3,M}
+ b_1 r^{2/3}ds^2_{6,E}, \\
e^{\phi} &= b_2 r^{-1},
\end{align}
where
\begin{align}
b_1 = \frac{3}{2^{7/3}} h^{2/3},
\qquad
b_2 = \frac{4}{3m} \sqrt{\frac{5}{3}}.
\end{align}
and $ds^2_{3,M}$ and $ds^2_{6,E}$ denote flat three-dimensional Minkowski and six-dimensional Euclidean metrics, respectively. This background satisfies the ten-dimensional dilaton and Einstein equations. 

\subsection{Completing the geometry with intersecting D4-brane stacks}

To obtain the full brane geometry, we introduce D4-branes dual to the unbounded $F_4$ fluxes. On a toroidal internal manifold these fluxes wrap three distinct two-tori, which corresponds to three intersecting stacks of D4-branes, as shown in Table~\ref{dgkt} (see also \cite{Apers:2022vfp,Kounnas:2007dd}).

\begin{table}[h!]
    \begin{center}
        \begin{tabular}{|c|c|c|c|c|c|c|c|c|c|c|}
            \hline
             & $t$ & $x^1$ & $x^2$ & $r$ & $z_1$ & $z_2$ & $z_3$ & $z_4$ & $z_5$ & $z_6$ \\
            \hline
            \textbf{D4} & $\otimes$ & $\otimes$ & $\otimes$ & & $\otimes$ & $\otimes$ & & & & \\
            \hline
            \textbf{D4} & $\otimes$ & $\otimes$ & $\otimes$ & & & & $\otimes$ & $\otimes$ & & \\
            \hline
            \textbf{D4} & $\otimes$ & $\otimes$ & $\otimes$ & & & & & & $\otimes$ & $\otimes$ \\
            \hline
        \end{tabular}
        \caption{D4-brane domain walls in the DGKT setup. The coordinates $(t,x^1,x^2)$ span $ds^2_{3,M}$ and the internal coordinates $z_i$ span $ds^2_{6,E}$.}
        \label{dgkt}
    \end{center}
\end{table}

The simplest way to incorporate these D4-branes is to associate a harmonic function $H(r)$ to each stack and apply the standard harmonic superposition rule \cite{Tseytlin:1996bh} for intersecting branes. We assume a factorised ansatz in which the D4-brane contributions multiply the background geometry obtained from flux backtracking. This leads to the following ten-dimensional string-frame metric:

\begin{tcolorbox}
\begin{aleq}\label{DGKT_full}
    ds_{10}^2 &= H(r)^{3/2}\, dr^2
    + r^{-10/9} H(r)^{-3/2}\, ds^2_{3,M}
    + b_1\, r^{2/3} H(r)^{1/2}\, ds^2_{6,E}, \\
    e^{\phi} &= b_2\, r^{-1} H(r)^{-3/4}.
\end{aleq}
\end{tcolorbox}

To check that this procedure is consistent, we substitute the ansatz into the ten-dimensional Einstein and dilaton equations. These are solved provided the harmonic function takes the form
\begin{tcolorbox}
\begin{equation}
    H(r) = c_1 + c_2\, r^{-4/3}.
\end{equation}
\end{tcolorbox}

In the next subsection we show that this geometry reproduces the DGKT vacua in the near-horizon limit.

This construction differs from that of \cite{Kounnas:2007dd} because we dualise only the fluxes that scale with $N$, rather than all fluxes. This cleanly separates a large-$N$ brane sector from a background supported by the remaining fluxes ($F_0$ and $H_3$), which dominate at large radial distance.

\subsection{Near-horizon limit and matching to DGKT scalings}

In the near-horizon region, the warp factor is dominated by
\begin{align}
    H(r) \sim c_2\, r^{-4/3},
\end{align}
and the ten-dimensional metric simplifies to
\begin{align}
    ds_{10}^2 
    &= \frac{dr^2}{r^2} + r^{8/9}\, ds_{3,M}^2 + ds_{6,E}^2 \\
    &= \frac{d\rho^2}{\rho^2} + \rho^2\, ds_{3,M}^2 + ds_{6,E}^2,
\end{align}
where $\rho = r^{4/9}$. 
This is exactly the metric for $\mathrm{AdS}_4 \times \text{torus}$ and in the same limit the dilaton approaches a constant \footnote{More precisely, the near-horizon limit consists of taking $\alpha'\rightarrow 0$ while keeping $U:=r^{4/9}/\alpha'$ constant.}

We now check whether this near-horizon solution reproduces the flux scalings of the DGKT vacua given in \eqref{DGKTprop1} and \eqref{DGKTprop2}. Reintroducing the flux dependence in the near-horizon region, the metric becomes
\begin{align}
    ds_{10}^2 &= c_2^{3/2}\, \frac{d\rho^2}{\rho^2} +  \rho^2\, ds_{3,M}^2 + c_2^{1/2}\, h^{2/3}\ ds_{6,E}^2, \\
    e^\phi &\sim c_2^{-3/4} m^{-1}.
\end{align}
Imposing flux quantisation for the D4-brane charge gives
\begin{align}
    N = \int_{\Sigma_4} F_4 
    = c_2\, h^{4/3} m,
\end{align}
where $\Sigma_4$ is the four-cycle wrapped by the $F_4$-flux. This determines the constant $c_2$ as
\begin{align}
    c_2 = \frac{N}{m h^{4/3}}.
\end{align}
Plugging this in, we find that the internal volume and string coupling scale as
\begin{align}
    \mathrm{vol}_S \sim N^{3/2} m^{-3/2}, 
    \qquad e^\phi \sim N^{-3/4} m^{-1/4} h,
\end{align}
and the central charge of the dual CFT behaves as
\begin{align}
    c \sim \left( \frac{L_{\mathrm{AdS}}}{l_p} \right)^2 \sim N^{9/2} m^{-5/2} h^{-4}.
\end{align}
These match exactly with the scalings of the original DGKT vacua, providing a strong consistency check of the flux backtracking approach.

We provide more examples of flux vacua where we apply this procedure in Appendix A.
\section{Decoupling of the DGKT brane theory}
A natural application of the full ten-dimensional metric \eqref{DGKT_full} is to check whether the brane sector decouples from the bulk.

\subsection{Gravitational redshift}
The metric \eqref{DGKT_full} converted to Einstein frame is given by,
\begin{align}
    ds_{10,E}^2 &= H(r)^{15/8} r^{1/2}\, dr^2 
    + r^{-11/18}\, H(r)^{-9/8}\, ds^2_{3,M} 
    + r^{7/6}\, H(r)^{7/8}\, ds^2_{6,E}, \\
    H(r) &= c_1 + c_2\, r^{-4/3}.
\end{align}
The non-compact coordinates are multiplied by a warp factor
\begin{aleq}
    f(r) = r^{-11/36}\, H(r)^{-9/16}.
\end{aleq}
As a consequence, an excitation of local energy $E_0$ near the branes, when observed by an asymptotic observer at large fixed radial position $r=R$, is redshifted according to
\begin{aleq}
    E_R = \lim_{r\rightarrow 0} \frac{f(r) E_0}{f(R)} \approx \lim_{r\rightarrow 0} R^{11/36} r^{4/9} E_0 =0,
\end{aleq}
which suggests that the branes decouple from the bulk.

\subsection{Infinite potential barrier}

To probe the decoupling of the D4-brane worldvolume theory more carefully, we study the propagation of graviton (and scalar) fluctuations in the background~\eqref{DGKT_full}. A closely related analysis was originally used to establish the decoupling of D3-branes in flat spacetime~\cite{Klebanov:1997kc, Gubser:1997yh}. There, it was shown that the absorption cross-section on the branes for bulk waves of intrinsic energy $\omega$, sent in from large radial distance, vanishes in the low-energy limit $\omega \to 0$.

Another way of seeing this is that the effective radial potential, given by
\begin{aleq}
    V_{D3}(r) =-\omega^2 \frac{r^4+L^4}{r^4} + \frac{15}{4r^2},
\end{aleq}
where $L$ is the AdS radius,
for such waves develops an infinitely high barrier near the branes, preventing low-energy modes from reaching the asymptotic region. We will follow this second method in this section.

We consider traceless transverse graviton perturbations of the form
\begin{align}
    h_{\mu\nu}(r,t) = e^{-i \omega t}\, h(r)\, \epsilon_{\mu\nu},
\end{align}
with polarization along the directions spanned by $ds_{3,M}^2$ in the background \eqref{DGKT_full}.

The linearized Einstein equations then reduce to
\begin{align}
&\Big(
81 \big(r^{4/3} + R^{4/3}\big)^{5} \omega^{2}
+ 4 r^{8/9} \big(55 r^{8/3} + 20 r^{4/3} R^{4/3} - 8 R^{8/3}\big)
\Big) h(r) \nonumber \\ &
+ 9 r^{17/9} \big(r^{4/3} + R^{4/3}\big)
\Big(
(41 r^{4/3} + 5 R^{4/3}) h'(r)
+ 9 r \big(r^{4/3} + R^{4/3}\big) h''(r)
\Big)
= 0.
\end{align}

\begin{figure}[h!]
    \centering
    \includegraphics[width=0.7\linewidth]{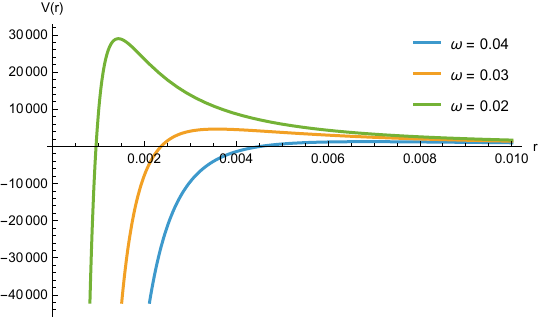}
    \caption{Effective potential $V(r)$ for different values of $\omega$.}
    \label{fig:s-wave-potential}
\end{figure}

We simplify this equation by redefining the fluctuation as
\begin{equation}
    h(r) = r^{-5/18} \big(r^{4/3} + R^{4/3}\big)^{-3/2}\, u(r),
\end{equation}
which brings the radial equation into a Schr\"odinger-like form,
\begin{equation}
    u''(r)
    - \left[
    -\omega^2 \frac{\big(r^{4/3} + R^{4/3}\big)^3}{r^{26/9}}
    + \frac{7}{36 r^2}
    \right] u(r) = 0.
\end{equation}
This allows us to identify an effective potential for the graviton modes,
\begin{tcolorbox}
\begin{equation}\label{DGKTpotbar}
V_{\text{DGKT}}(r)
= -\omega^2 \frac{\big(r^{4/3} + R^{4/3}\big)^3}{r^{26/9}}
+ \frac{7}{36 r^2}.
\end{equation}
\end{tcolorbox}
The same potential is obtained for minimal $s$-wave scalar fluctuations. Figure~\ref{fig:s-wave-potential} shows $V(r)$ for several values of~$\omega$.

In the low-energy limit $\omega \to 0$, the potential develops an infinitely high barrier near $r = 0$. As a result, perturbations localised close to the branes decay exponentially toward larger radial distances, signaling decoupling from the bulk. We also note that, unlike in standard Freund--Rubin backgrounds, the potential does not approach a constant at large $r$, but instead approaches negative infinity.

\subsection{Discussion}

In summary, we have found that
\begin{enumerate}
    \item Excitations near the D4-brane worldvolumes are infinitely redshifted relative to observers at a fixed radial position.
    \item Near the branes, there is an infinite potential barrier for waves in the decoupling limit.
\end{enumerate}
These results indicate that the D4-brane worldvolume theory successfully decouples from the bulk.

Recently, however, it was argued in \cite{Bedroya:2025ltj} (see also \cite{Bedroya:2025fie}) that brane theories underlying scale-separated AdS vacua cannot decouple, based on an analysis of the asymptotic region $r \to \infty$.

We can analyse this asymptotic region by setting $H(r)\rightarrow 1$ in the full ten-dimensional geometry \eqref{DGKT_full}, and find
\begin{aleq}\label{as}
ds_{10}^2 &= dr^2
+ r^{-10/9} ds^2_{3,M}
+r^{2/3}ds^2_{6,E}, \qquad
e^{\phi} \sim   r^{-1}.
\end{aleq}
This coincides precisely with the background obtained via flux backtracking with Romans mass and $H_3$ flux in \eqref{BG}. Although the flux-backtracking method was introduced to identify the singular geometry probed by branes, it also determines the asymptotic region, since the $F_4$ flux sourcing the D4-branes becomes subdominant at large internal volume. Upon dimensional reduction, the resulting four-dimensional metric
\begin{aleq}
    ds_4^2 = dr^2 + r^{26/27} ds^2_{3,M}
\end{aleq}
agrees with the asymptotic geometry described in \cite{Bedroya:2025ltj}. In that work, the brane geometry is constructed by solving the asymptotic flow equations using a dynamical systems approach and postulating a matching onto an AdS throat. The explicit solution \eqref{DGKT_full} obtained here provides a concrete ten-dimensional interpolation between the AdS throat and the asymptotic region.

From \eqref{as}, we read the warp factor for the non-compact directions
\begin{aleq}\label{blue}
    f_{\text{asymptotic}}(r) = r^{-5/9} M_s^{-1}.
\end{aleq}
This implies that the DGKT vacua do not satisfy the decoupling condition proposed in \cite{Bedroya:2025ltj},
\begin{aleq}
    \lim_{r\to\infty} \frac{M_p}{\Lambda_s f(r)} < \infty,
\end{aleq}
where the species scale $\Lambda_s$ is identified with the string scale $M_s$. The quantity $g\equiv M_p\Lambda_s^{-1} f(r)^{-1}$ is interpreted in \cite{Bedroya:2025ltj} as the coupling of perturbations to gravity. If this diverges, then perturbations from the branes are argued not to decouple from the bulk. However, this conclusion assumes that brane-localised perturbations can reach the asymptotic region, despite the presence of an infinite potential barrier.

\begin{figure}[h!]
    \centering
    \begin{subfigure}[t]{0.45\linewidth}
        \centering
        \includegraphics[width=\linewidth]{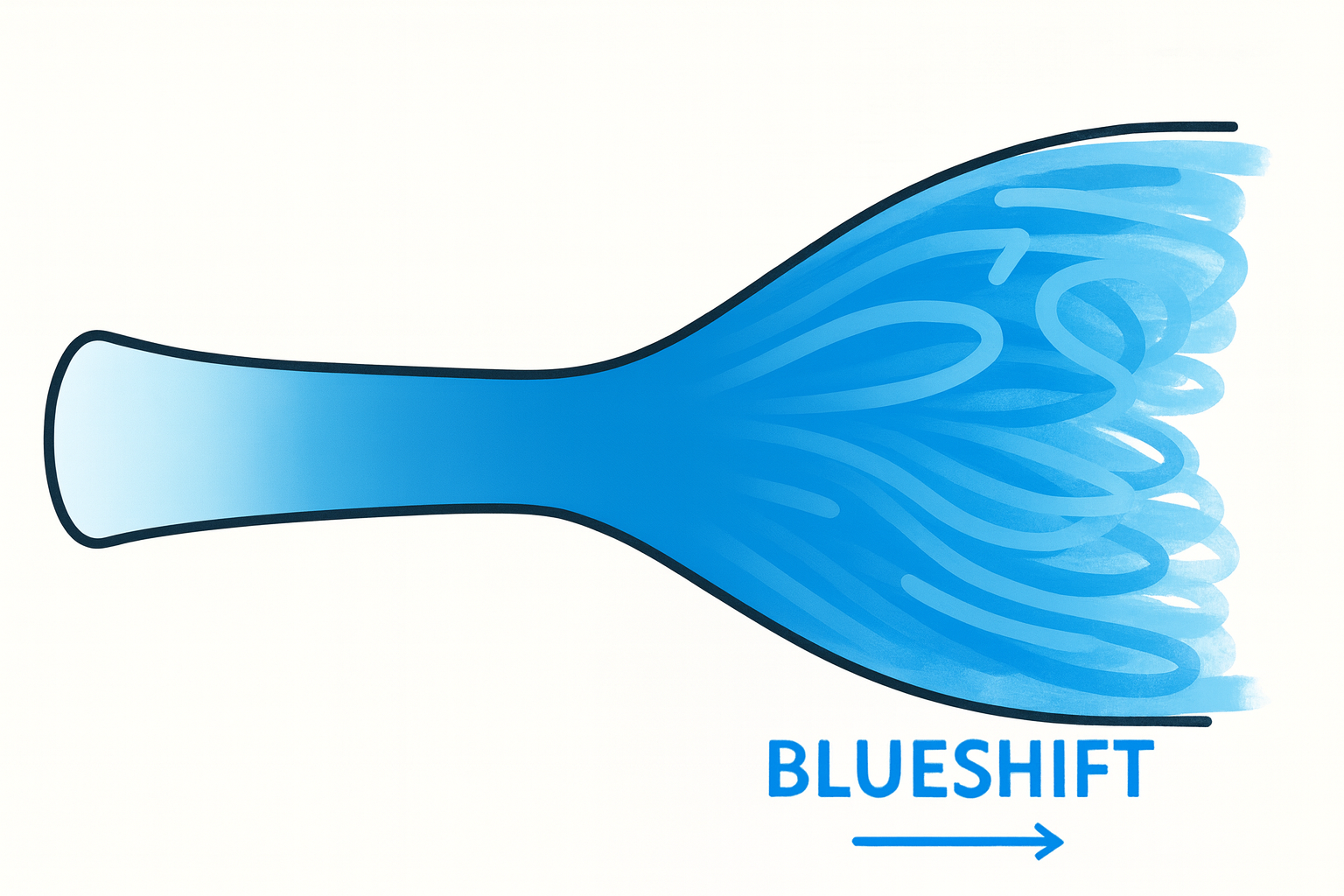}
        \label{fig:blueshift}
    \end{subfigure}
    \hfill
    \begin{subfigure}[t]{0.47\linewidth}
        \centering
        \includegraphics[width=\linewidth]{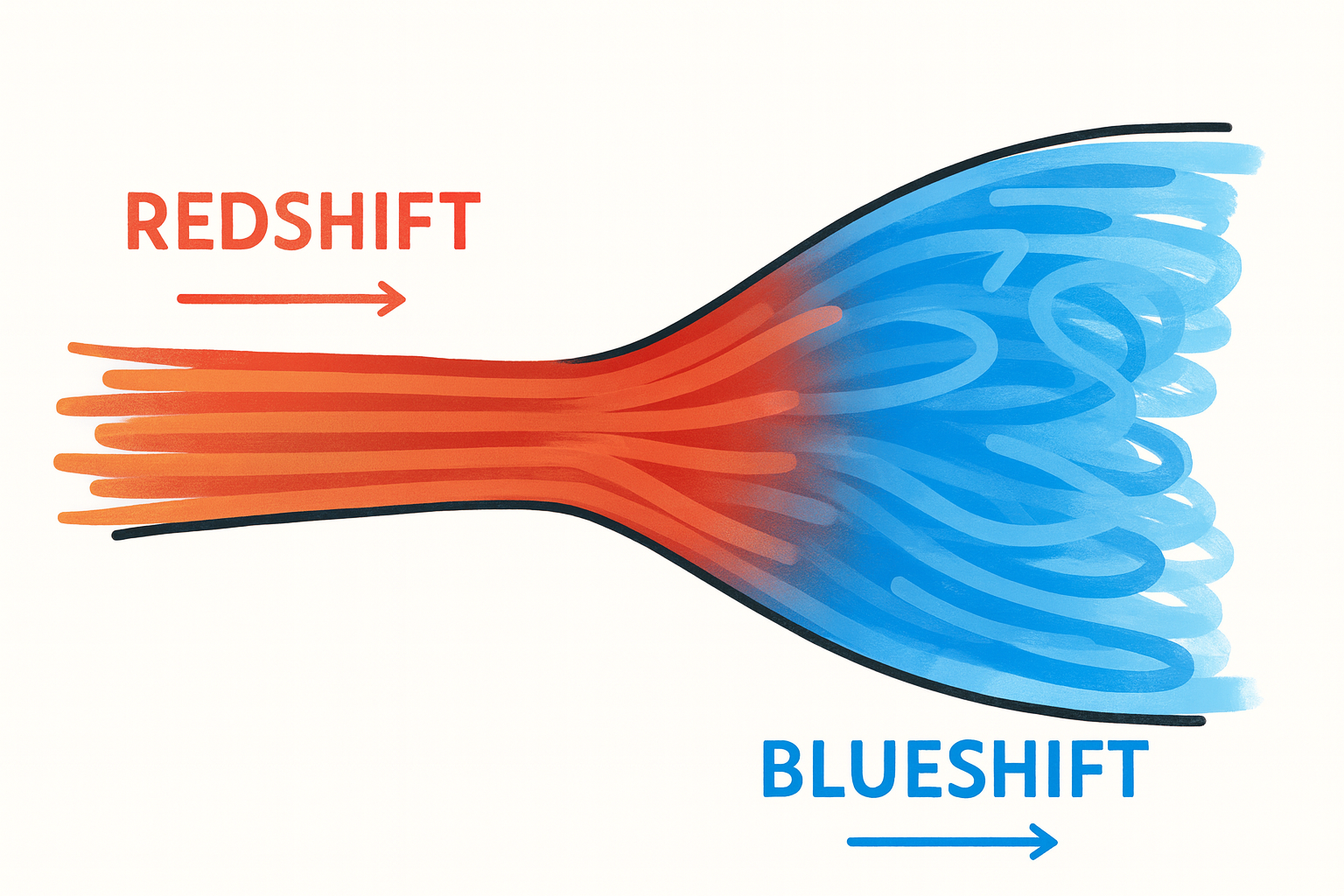}
        \label{fig:redblueshift}
    \end{subfigure}
    \caption{The analysis of \cite{Bedroya:2025ltj} only considers blueshift in the asymptotic region (left). We point out that there is opposite redshifting behaviour in the AdS throat (right).}
    \label{fig:shifts}
\end{figure}

From a gravitational redshift perspective \eqref{blue} implies that modes localised at a fixed radial position become increasingly blueshifted relative to the string scale when measured by observers at increasing radial distances. Once the D4-brane stacks are included, the metric is modified by the associated harmonic functions. Inside the AdS throat, the scale factor is instead approximately
\begin{aleq}
    f_{\text{throat}}(r) = H^{-3/4}(r)\, r^{-5/9} M_s^{-1} \approx r^{4/9}\, M_s^{-1},
\end{aleq}
which has the opposite behaviour to the asymptotic region. As a result, modes localised near the branes at the bottom of the throat appear increasingly redshifted to any observer at fixed radial distance. These two qualitatively different behaviours are illustrated schematically in Figure~\ref{fig:shifts}.

The blueshift in the asymptotic region is very non-standard, and it is an interesting
open question whether this leads to any pathology in a holographic description. In standard AdS/CFT examples, the branes decouple from weakly coupled gravity at infinity, whereas it is unclear whether the DGKT asymptotic region \eqref{as} admits a controlled description. We have two remarks.

\begin{itemize}
\item 
The asymptotic metric~\eqref{BG} has a vanishing dilaton, $e^{\phi}\to 0$, together with vanishing curvature invariants in the radial asymptotic region,
\begin{equation}
R = -\frac{28}{27}\,r^{-2}, 
\qquad 
R_{\mu\nu}R^{\mu\nu} = \frac{1672}{729}\,r^{-4}, 
\qquad 
R_{\mu\nu\rho\sigma}R^{\mu\nu\rho\sigma} = \frac{31712}{2187}\,r^{-4}.
\end{equation}
As a consequence, both $\alpha'$ and higher-derivative corrections are suppressed, and the ten-dimensional supergravity description becomes increasingly reliable. In contrast, any effective four-dimensional description breaks down in this regime. From this perspective, string theory remains well-defined on the asymptotic background.

\item 
The asymptotic behaviour is closely related to the presence of a nonzero Romans mass. Explicit examples of AdS$_4$ vacua in massive IIA with proposed holographic duals include the Romans-mass deformed ABJM solutions~\cite{Gaiotto:2009mv} and the $S^6$ consistent truncations of massive IIA~\cite{Guarino:2015jca}. The latter are dual to three-dimensional Chern--Simons--matter theories, for which a precision match has been achieved between the gravitational free energy (scaling as $N^{5/3} m^{1/3}$, with $m$ the Romans mass) and the field-theoretic partition function. The brane construction consists of $N$ D2-branes deformed by the Romans mass.

Using the flux-backtracking approach, one can determine the asymptotic geometry far from the brane sources. In this regime, the scalar potential is dominated by the internal curvature of the $S^6$ and the Romans mass. A solution to the domain-wall equations sourced by these contributions was obtained in~\cite{Apers:2025pon},
\begin{equation}
u(r) \sim r^{2/5}, 
\qquad 
s(r) \sim r^{4/5}, 
\qquad 
A(r) = \frac{19}{25}\ln r,
\end{equation}
where $u=\mathrm{vol}_S^{1/3}$ denotes the volume modulus, $e^\phi$ is the four-dimensional dilaton, and $A(r)$ is the warp factor. This corresponds to the ten-dimensional geometry
\begin{equation}
ds_{10}^2 = dr^2 + r^{-2/5} ds_{3,M}^2 + r^2 ds_{6,E}^2,
\qquad 
e^\phi \sim r^{-1}.
\end{equation}
The scale factor for the non-compact directions,
\begin{equation}
f_{\text{asymptotic}}(r) = r^{-1/5} M_s^{-1},
\end{equation}
grows more slowly than the string length at large radius. As a result,
\begin{equation}
\lim_{r\to\infty} \frac{M_p}{\Lambda_s f(r)} = \infty.
\end{equation}
This example suggests that such non-standard asymptotic behaviour does not necessarily rule out the existence of a well-defined gauge-theory dual.
\end{itemize}

\subsection{Conclusion}
The DGKT brane geometry has a peculiar asymptotic region where perturbations appear increasingly blueshifted to observers at larger and larger radial distances. However, because of the redshift in the AdS throat, perturbations near the branes will appear infinitely redshifted to any observer at fixed radial distance.

Regardless of whether the asymptotic region is under control, the region near the branes is controlled and therefore the presence of an infinite potential barrier there in the decoupling limit is unambiguous, implying a decoupling from the bulk from this perspective.

\section*{Acknowledgments}
I am grateful to Miguel Montero and Irene Valenzuela for helpful discussions and for comments on an earlier draft. I also thank Miquel Aparici, Ivano Basile, Alek Bedroya, Joseph Conlon, Ludo Fraser-Taliente, Dieter Lüst, Antonia Paraskevopoulou, Alejandro Puga, Muthusamy Rajaguru, Alessandro Tomasiello, Vincent Vanhemelryck, and Thomas Van Riet for interesting conversations related to this work. 

I am supported by the ERC Starting Grant QGuide101042568 - StG 2021. This work was initiated in part at the Aspen Center for Physics, which is supported by a grant from the Simons Foundation (1161654, Troyer).

\appendix

\section{More examples of the flux backtracking method}

In Section~1, we reconstructed the DGKT brane geometry using the flux backtracking method with bounded fluxes, and introduced the unbounded fluxes through harmonic functions for the dual branes, placed on the background in a factorised manner. Here we give a few additional examples to illustrate that this construction works more generally.

The first example concerns orbifolds of Freund--Rubin vacua, such as
$\mathrm{AdS}_5 \times S^5/\mathbb{Z}_k$, where we explicitly recover the
$k$-dependence of the solution from the near-horizon geometry. The second
example is the ABJM construction in ten-dimensional type IIA language,
leading to a geometry of D2-branes probing a singularity. Upon uplifting
to eleven dimensions, this is interpreted as M2-branes on
$\mathbb{C}^4/\mathbb{Z}_k$.

Finally, we discuss scale-separated AdS$_4$ vacua in type IIA without Romans mass, which are related by two T-dualities to the DGKT vacua \cite{Cribiori:2021djm}. These correspond to compactifications on an Iwasawa manifold in the presence of $F_6$ and $F_2$ fluxes. The $F_6$ flux is unbounded and can be dualised to D2-branes. Some legs of the $F_2$ flux are unbounded as well and can be dualised to stacks of D6-branes. The flux backtracking procedure then offers different perspectives. One option is to solve the flow equations with all $F_2$ fluxes included, yielding a background probed by D2-branes that closely resembles the ABJM background. Alternatively, one may dualise all unbounded fluxes to branes and study the background probed by a D2--D6 system.

For completeness, we also perform a decoupling analysis for the D-brane theories in ABJM and in these scale-separated vacua, computing an effective potential for scalar $s$-waves in the corresponding backgrounds and verifying the presence of an infinite potential barrier.

\subsection{Orbifolds of Freund-Rubin vacua}
\subsubsection*{AdS$_5 \times S^5/\mathbb{Z}_k$ vacua}
The effective scalar potential for the volume modulus $\mathrm{vol}_S$ with contributions from the $F_5$ flux and curvature takes the schematic form,
\begin{aleq}
    V = N^2 \mathrm{vol}_S^{-8/3} -k^{-2/5} \mathrm{vol}_S^{-16/5}.
\end{aleq}
The flow equations sourced by the flux term are solved by \cite{Apers:2025pon}
\begin{aleq}
    \mathrm{vol}_S(r) \sim k^{-3/8} r^{15/8}, \quad ds_5^2 = dr^2 + e^{2A(r)} ds_{4,M}^2 = dr^2 + r^{5/4}  ds_{4,M}^2,
\end{aleq}
where $ds_{4,M}^2$ is a four-dimensional Minkowski metric
and the uplift to ten dimensions of this metric is,
\begin{aleq}
    ds_{10}^2 = dr^2 +  ds_{4,M}^2 + r^2 k^{-2/5} ds_5^2,
\end{aleq}
where $ds_5^2$ is the metric on the internal manifold,
which is the background probed by D3-branes. Placing a stack of $N$ D3-branes in this background modifies the geometry to
\begin{aleq}
    ds_{10}^2 = H(r)^{\frac{1}{2}} dr^2 
    + H(r)^{-\frac{1}{2}} ds_{4,M}^2 
    + r^2 k^{-2/5} H(r)^{\frac{1}{2}} ds_5^2,
\end{aleq}
with harmonic function
\begin{aleq}
    H(r) = c_1 + c_2 r^{-4}.
\end{aleq}

The constant $c_2$ is fixed by flux quantisation,
\begin{aleq}
    N = \frac{1}{(2\pi)^4 g_s \alpha'^2} \int_{S^5/\mathbb{Z}_k} F_5,
\end{aleq}
where
\begin{aleq}
    \int_{S^5/\mathbb{Z}_k} F_5 
    &= \int_{S^5/\mathbb{Z}_k} \partial_r H^{-1} \, r^5 \, \mathrm{vol}(S^5/\mathbb{Z}_k) \\
    &= 4c_2 \frac{\mathrm{vol}(S^5)}{k} 
    = 4c_2 \frac{\pi^3}{k}.
\end{aleq}
This gives
\begin{aleq}
    c_2 = \frac{(2\pi)^4 g_s \alpha'^2 N k}{4\pi^3} 
    = 4\pi g_s \alpha'^2 N k.
\end{aleq}

In the near-horizon limit, the metric becomes
\begin{aleq}
    ds_{10}^2
    &= \frac{r^2}{\sqrt{Nk}}\, ds_{4,M}^2 
    + \sqrt{Nk}\, \frac{dr^2}{r^2} 
    + \sqrt{N}\, k^{1/10}\, ds_5^2,
\end{aleq}
which is precisely AdS$_5 \times S^5/\mathbb{Z}_k$. The internal volume in string units is
\begin{aleq}
    \mathrm{vol}(S^5/\mathbb{Z}_k) \sim N^{5/4} k^{1/4}.
\end{aleq}

From the metric, we extract the AdS radius
\begin{aleq}
    \frac{R_{\text{AdS}}}{l_s} = N^{1/4} k^{1/4},
\end{aleq}
and, using $l_s \sim l_p \cdot \left(N^{5/12} k^{1/12}\right)$, we find that the central charge scales as
\begin{aleq}
    c \sim N^2 k,
\end{aleq}
in agreement with expectations.

\subsubsection*{AdS$_4 \times S^7/\mathbb{Z}_k$ vacua}
In an analogous way as above, we obtain the following singular metric which is probed by a stack of M2-branes,
\begin{aleq}\label{FRmetric4}
    ds_{11}^2 = H(r)^{\frac{1}{3}}dr^2 + H(r)^{-\frac{2}{3}}ds_{3,M}^2 + r^2 k^{-2/7} ds_7^2
\end{aleq}
with 
\begin{aleq}
   H(r) = c_1 + c_2r^{-6}.
\end{aleq}
Imposing flux quantisation, we learn that
\begin{aleq}
    c_2 = 2^5 \pi^2 l_p^6 Nk.
\end{aleq}
Therefore we can deduce the following flux scalings from the near-horizon geometry,
\begin{aleq}
    \mathrm{vol}(S^7/\mathbb{Z}_k) \sim N^{7/6} k^{1/6},
\end{aleq}
and
\begin{aleq}
    \frac{R_{AdS}}{l_{11}} = N^{1/6} k^{1/6}, \quad c =N^{3/2} k^{1/2},
\end{aleq}
as required.

\subsubsection*{AdS$_7 \times S^4/\mathbb{Z}_k$ vacua}
In exactly the same way, the metric probed by M5-branes is
\begin{aleq}\label{FRmetric7}
    ds_{11}^2 = H(r)^{\frac{2}{3}}dr^2 + H(r)^{-\frac{1}{3}}ds_{6,M}^2 + r^2 k^{-1/2} ds_4^2,
\end{aleq}
with 
\begin{aleq}
     H(r) = c_1 + c_2r^{-3}, \quad c_2 = \pi l_p^3 Nk.
\end{aleq}
The near-horizon geometry has the following properties.
\begin{aleq}
     \mathrm{vol}(S^4/\mathbb{Z}_k)  \sim N^{4/3} k^{1/3}, \quad l_4 \sim l_{11} N^{4/15}k^{1/15}.
\end{aleq}
The AdS radius and the central charge then equal
\begin{aleq}
    \frac{R_{AdS}}{l_{11}} = N^{1/3} k^{1/3}, \quad c =N^{3} k^{2}.
\end{aleq}
as required.

\subsection{ABJM vacua: ten-dimensional perspective}
To find the geometry probed by D2-branes we perform the backtracking procedure with $k$ units of $F_2$ flux and internal curvature and obtain the metric and dilaton \cite{Apers:2025pon}
\begin{aleq}
    ds_{10}^2 
    = dr^2 
    + r^{2/3} k^{-1/2} ds_{3,M}^2 
    + r^2 ds_6^2,
    \qquad
    e^{\phi} \sim r\, k^{-1},
\end{aleq}
up to numerical prefactors.
Placing D2-branes in this background leads to the metric
\begin{aleq}\label{ABJM}
    ds_{10}^2 
    = H(r)^{1/2} dr^2
    + r^{2/3} k^{-1/2} H(r)^{-1/2} ds_{3,M}^2 
    + r^2 H(r)^{1/2} ds_6^2,
\end{aleq}
with dilaton
\begin{aleq}
    e^{\phi} = r\, k^{-1} H(r)^{1/4}.
\end{aleq}
The dilaton equation implies
\begin{aleq}
    H''(r) r + 5 H'(r) = 0,
\end{aleq}
which is solved by
\begin{aleq}
    H(r) = c_1 + \frac{c_2}{r^4}.
\end{aleq}
Flux quantisation determines the scaling of $c_2$. Up to numerical factors,
\begin{aleq}
    N \sim \int F_6 \sim c_2\, k,
\end{aleq}
so that
\begin{aleq}
    c_2 \sim \frac{N}{k}.
\end{aleq}
It follows that the internal volume and dilaton scale as
\begin{aleq}
    \mathrm{vol}_S \sim c_2^{3/2} \sim N^{3/2} k^{-3/2},
    \qquad
    e^{\phi} \sim N^{1/4} k^{-5/4}.
\end{aleq}
The AdS length scale is given by
\begin{aleq}
    \ell_{\mathrm{AdS}} \sim c_2^{1/4} l_s \sim N^{1/4} k^{-1/4} l_s,
\end{aleq}
which correctly reproduces the expected $N$- and $k$-scaling of the central charge:
\begin{aleq}
    c \sim \ell_{\mathrm{AdS}} ^2 l_p^{-2} \sim N^{3/2} k^{1/2}.
\end{aleq}

\subsubsection*{Decoupling of the D2-brane theory}

In this case, the non-compact directions in \eqref{ABJM} are multiplied by a radial function
\begin{aleq}
    k^{-1/2} f(r) = r^{2/3} H(r)^{-1/2},
\end{aleq}
which scales as $f(r) \sim r^{2/3}$ in the asymptotic region and as $f(r) \sim r^{8/3}$ in the near-horizon region. As a result, modes seen by asymptotic observers are strongly redshifted.

The effective potential for scalar $s$-waves in this background takes the form
\begin{aleq}
    V_{\text{eff}}(r) = \frac{15}{4 r^2} 
    - \frac{\omega^2 \left(r^4 + R^4\right)}{r^{14/3}},
\end{aleq}
where we have set $c_1 = 1$ and $c_2 = R$. This potential has a positive barrier, which becomes infinitely high in the limit $\omega \to 0$.

\subsection{Scale-separated AdS$_4$ vacua in massless IIA}

The superpotential takes the schematic form
\begin{aleq}
    W = N + k_1 T_1 T_2 + k_2 T_1 T_3 + k_3 T_2 T_3 + S T_1 .
\end{aleq}
Here $F_6 \sim N$ and there are three different legs of $F_2$ flux $\sim k_i$, where $k_3$ is bounded. The $T_i$ are the complexified K\"ahler moduli (with real parts $u_i$), and $S$ is the complexified four-dimensional dilaton.

\subsubsection*{Perspective 1: Background probed by D2-branes}

The first option is to perform flux backtracking with all $F_2$ fluxes turned on,
\begin{aleq}
    W_{\text{res}} = k_1 T_1 T_2 + k_2 T_1 T_3 + k_3 T_2 T_3 + S T_1 ,
\end{aleq}
in which case we find the background
\begin{aleq}
    ds_{10}^2 
    &= dr^2 
    + r^{2/3} k_3^{-1/2} ds_{3,M}^2 
    + r^2 \left(
        \frac{k_3^2}{k_1 k_2} dy_{1,2}^2
        + \frac{k_3}{k_1} dy_{3,4}^2
        + \frac{k_3}{k_2} dy_{5,6}^2
    \right),
\end{aleq}
where $y_i$ parametrise the internal Iwasawa manifold. The dilaton is
\begin{aleq}
    e^{\phi} = r\, \frac{k_3}{k_1 k_2}.
\end{aleq}

Placing D2-branes in this background leads to
\begin{aleq}\label{mDGKT}
    ds_{10}^2 
    &= H(r)^{1/2} dr^2 
    + r^{2/3} k_3^{-1/2} H(r)^{-1/2} ds_{3,M}^2 \nonumber\\
    &\quad
    + r^2 H(r)^{1/2}
    \left(
        \frac{k_3^2}{k_1 k_2} dy_{1,2}^2
        + \frac{k_3}{k_1} dy_{3,4}^2
        + \frac{k_3}{k_2} dy_{5,6}^2
    \right),
\end{aleq}
and
\begin{aleq}
    e^{\phi} \sim r\, \frac{k_3}{k_1 k_2} H(r)^{1/4}.
\end{aleq}

Choosing the harmonic function
\begin{aleq}
    H(r) = c_1 + c_2 r^{-4},
\end{aleq}
the dilaton and internal volume become constant in the near-horizon region.

Flux quantisation gives, up to numerical factors,
\begin{aleq}
    N \sim c_2 \frac{k_3^3}{k_1 k_2},
\end{aleq}
so that
\begin{aleq}
    c_2 \sim \frac{N k_1 k_2}{k_3^3}.
\end{aleq}

It follows that the dilaton and internal moduli scale as
\begin{aleq}
    e^{\phi} 
    &\sim N^{1/4} k_3^{1/4} k_1^{-3/4} k_2^{-3/4}, \\
    u_1 &\sim \sqrt{\frac{N k_3}{k_1 k_2}}, 
    \qquad
    u_2 \sim \sqrt{\frac{N k_1}{k_2 k_3}}, 
    \qquad
    u_3 \sim \sqrt{\frac{N k_2}{k_1 k_3}} .
\end{aleq}
The internal volume scales as
\begin{aleq}
    \mathrm{vol}_S \sim \sqrt{\frac{N^3}{k_1 k_2 k_3}} .
\end{aleq}

The string length is then
\begin{aleq}
    l_s \sim N^{1/2} \sqrt{\frac{k_1 k_2}{k_3}} .
\end{aleq}
From the AdS warp factor,
\begin{aleq}
    \frac{l_{\mathrm{AdS}}}{l_s} \sim c_2^{1/4}
    \sim \frac{N^{1/4}(k_1 k_2)^{1/4}}{k_3^{3/4}},
\end{aleq}
and therefore
\begin{aleq}
    \frac{l_{\mathrm{AdS}}}{l_{p,4}} 
    \sim \frac{N^{3/4}(k_1 k_2)^{3/4}}{k_3^{5/4}} .
\end{aleq}
These scalings agree with those derived from the scalar potential.

\subsubsection*{Perspective 2: Background probed by a D2--D6 system}

Alternatively, we only keep the bounded $F_2$ flux and the curvature term in the superpotential,
\begin{aleq}
    W_{\text{res}} = k_3 T_2 T_3 + S T_1 .
\end{aleq}
The resulting metric and dilaton are
\begin{aleq}
    ds_{10}^2 
    = dr^2 
    + k_3^{-1} r^{-10/9} ds_{3,M}^2 
    + k_3^2 r^{-2/3} dy_{1,2}^2
    + k_3 r^{-2/3} dy_{3,4}^2
    + k_3 r^{-2/3} dy_{5,6}^2 ,
\end{aleq}
\begin{aleq}
    e^\phi = k_3 r^{-5/3}.
\end{aleq}

\begin{table}[h!]
    \begin{center}
        \begin{tabular}{|c|c|c|c|c|c|c|c|c|c|c|}
            \hline
             & $t$ & $x^1$ & $x^2$ & $r$ & $y_1$ & $y_2$ & $y_3$ & $y_4$ & $y_5$ & $y_6$ \\
            \hline
            \textbf{D2} 
            & $\otimes$ & $\otimes$ & $\otimes$ & & & & & & & \\
            \hline
            \textbf{D6} 
            & $\otimes$ & $\otimes$ & $\otimes$ & & $\otimes$ & $\otimes$ & $\otimes$ & $\otimes$ & & \\
            \hline
            \textbf{D6} 
            & $\otimes$ & $\otimes$ & $\otimes$ & & $\otimes$ & $\otimes$ & & & $\otimes$ & $\otimes$ \\
            \hline
        \end{tabular}
        \caption{D2--D6 brane domain walls.}
        \label{TTdgkt}
    \end{center}
\end{table}

Next, we place a D2--D6 configuration (as shown in Table \ref{TTdgkt}) on this background. For the D2-branes we assign a harmonic function $H$, and for the D6-branes a harmonic function $K$:
\begin{aleq}
    ds_{10}^2 
    &= H^{1/2} K\, dr^2 
    + H^{-1/2} K^{-1} k_3^{-1} r^{-10/9} ds_{3,M}^2  \nonumber\\
    &\quad
    + H^{1/2} K^{-1} k_3^2 r^{-2/3} dy_{1,2}^2
    + H^{1/2} k_3 r^{-2/3} dy_{3,4}^2
    + H^{1/2} k_3 r^{-2/3} dy_{5,6}^2 ,
\end{aleq}
with dilaton
\begin{aleq}
    e^\phi = H^{1/4} K^{-3/2} k_3 r^{-5/3}.
\end{aleq}
For
\begin{aleq}
    H(r) \sim K(r) \sim r^{-4/3},
\end{aleq}
the dilaton and volume moduli approach constants in the near-horizon region, and an AdS vacuum is obtained.
\subsubsection*{Decoupling of D2-branes and the D2--D6 system}

Since the radial dependence of the metric and dilaton in \eqref{mDGKT} is identical to that of the ABJM background \eqref{ABJM}, the decoupling analysis proceeds in the same way.

For the D2--D6 system, the radial function multiplying the non-compact coordinates scales as
\begin{aleq}
    k_3^{-1} f(r) = H^{-1/2} K^{-1} r^{-10/9},
\end{aleq}
and exhibits the same asymptotic and near-horizon behaviour as the background probed by the triple D4-brane system in DGKT \eqref{DGKT_full}. The effective potential for scalar $s$-waves coincides with that of DGKT \eqref{DGKTpotbar}.

\bibliographystyle{utphys}
\bibliography{references}

\end{document}